# Seasonality can induce coexistence of multiple bet-hedging strategies in *Dictyostelium discoideum* via storage effect.

Ricardo Martínez-García[1],

Corina E. Tarnita[1,*].

1. Department of Ecology and Evolutionary Biology, Princeton University. Princeton NJ, 08544.

*Corresponding author; e-mail: ctarnita@princeton.edu

# Abstract

The social amoeba *Dictyostelium discoideum* has been recently suggested as an example of bet-hedging in microbes. In the presence of resources, amoebae reproduce as unicellular organisms. Resource depletion, however, leads to a starvation phase in which the population splits between aggregators, which form a fruiting body made of a stalk and resistant spores, and non-aggregators, which remain as vegetative cells. Spores are favored when starvation periods are long, but vegetative cells can exploit resources in environments where food replenishes quickly. The investment in aggregators versus non-aggregators can therefore be understood as a bet-hedging strategy that evolves in response to stochastic starvation times. A genotype (or strategy) is defined by the balance between each type of cells. In this framework, if the ecological conditions on a patch are defined in terms of the mean starvation time (i.e. time between the onset of starvation and the arrival of a new food pulse), a single genotype dominates each environment, which is inconsistent with the huge genetic diversity observed in nature. Here we investigate whether seasonality, represented by a periodic, wet-dry alternation in the mean starvation times, allows the coexistence of several strategies in a single patch. We study this question in a non-spatial (well-mixed) setting in which different strains compete for a common pool of resources over a sequence of growth-starvation cycles. We find that seasonality induces a temporal storage effect that can promote the stable coexistence of multiple genotypes. Two conditions need to be met in our model. First, there has to be a temporal niche partitioning (two well-differentiated habitats within the year), which requires not only different mean starvation times between seasons but also low variance within each season. Second, each season's well-adapted strain has to grow and create a large enough population that permits its survival during the subsequent unfavorable season, which requires the number of growth-starvation cycles within each season to be sufficiently large. These conditions allow the coexistence of two bet-hedging strategies. Additional tradeoffs among life-history traits can expand the range of coexistence and increase the number of coexisting strategies, contributing towards explaining the genetic diversity observed in *D. discoideum*. Although focused on this cellular slime mold, our results are general and may be easily extended to other microbes.

# 1. Introduction.

In *Dictyostelium discoideum* starvation triggers the aggregation of free-living amoebae and the development of a multicellular fruiting body made of reproductive spores and dead stalk cells. During aggregation, cells do not completely exclude genetic non-relatives and therefore chimeric fruiting bodies (made of at least two genotypes) can be formed. Fitness in *D. discoideum* has been traditionally equated to the number of spores [1]; in lab experiments, this has led to the establishment of a linear hierarchy of genotypes (or strains) that reflects the overrepresentation of certain genotypes in the spores of chimeras [2]. This result points to a decrease in the number of existing strains, which is incompatible with the huge diversity observed in natural isolates of *D. discoideum* [3]. Recent studies have suggested, however, that this inconsistency arises from the one-to-one correspondence between spore number and fitness, which is likely incomplete since it ignores various other fitness components, such as spore viability [4], and the role of vegetative non-aggregated cells [5], [6]. The existence of several tradeoffs among these components turns the fitness of *D. discoideum* into a more complex quantity [4], [7]. However, even in this more comprehensive framework coexistence remains puzzling and additional mechanisms need to be considered.

For the purpose of this study, since mutation rates in *D. discoideum* are very low [8], we consider that strategies cannot mutate into each other, and thus multi-strain coexistence can be studied within the well-established theoretical framework of species coexistence. Classic results from community ecology established that only one species can survive in communities where different species compete for one common resource [9]–[11]. Coexistence requires that different species be heterogeneous in the way they respond to and affect their biotic and abiotic environment [12]–[15]. *D. discoideum* presents, however, one additional issue: the different strains are hypothesized to hedge their bets in response to uncertain environmental conditions. Upon starvation, amoebae diversify their commitment in the formation of the fruiting body. As a result, some of them remain as non-aggregating vegetative cells. According to some models this population partitioning could represent a risk-spreading reproductive strategy [5], [6]. Non-aggregators readily start reproducing after resource replenishment, but they are less resistant to starvation. Spores survive longer starvation times but pay a cost when food recovers fast due to time-consuming fruiting body development and delayed germination. In this theoretical framework, how much a strain invests in each type of cell (aggregating versus non-aggregating) is a bet-hedging trait that evolves in response to uncertainty in the starvation times: slower-recovering environments, characterized by longer mean starvation periods, select for more spores; faster-recovering environments, characterized by shorter starvation periods, select for higher investments in non-aggregators [5], [6].

Spatial heterogeneity as a promoter of coexistence has recently been studied in *D. discoideum* [5] and low-to-moderate dispersal between multiple patches has been theoretically shown to allow the coexistence of several *D. discoideum* strains. Here we explore the role of temporal heterogeneity (in this case seasonality) in fostering coexistence. This turns out to be challenging because bet-hedging strategies are plastic, which allows them to average across different ecological conditions by reducing the variance of the fitness in order to minimize the risks of complete reproductive failure. This plasticity comes at the expense of diminishing the (arithmetic) mean fitness, since some offspring are always maladapted to a subset of environmental conditions [16]–[20]. In the presence of seasonality, which introduces a second characteristic scale in the environment by periodically switching its statistical properties, the optimal bet-hedging strategy may change. Hence, it is hard to discern when seasonality will cause the evolution of a new optimal bet-hedging strategy that averages over both seasons, and when it will lead to the coexistence of two season-specialist bet-hedging strategies (i.e. temporal niche partitioning [21]–[25]). To explore the necessary conditions that lead to a temporal niche partition and the coexistence of multiple bet-hedging strategies, we depart from previous studies that fix the environmental conditions during good and bad years, we explore the differences in the response of the species to that alternation [26], and we extend them by allowing both the statistical properties of the seasons and the responses of the species to vary.

Temporal coexistence of bet-hedging species has been studied both theoretically [12], [27] and experimentally [28] in seed banks of annual plants, one of the classic and best studied instances of bet-hedging [29]–[31]. According to previous studies, periodic changes in the ecological conditions during the year mediate the coexistence of several strains through a temporal storage effect [12], [25], [32] if three general requirements are satisfied: (i) changes in the environmental conditions favor different species (temporal niche partition), (ii) the rate at which populations decline together with the temporal scale of the environmental fluctuations avoids the extinction of non-favored species and (iii) the covariance between environment and competition intensity is opposite for high density and low density species, which allows species to have positive growth rates when they become less abundant and recover larger population sizes [26], [33]–[35].

In summary, here we aim to: (i) provide a theoretical starting point to unveil the ecological conditions that promote coexistence of microbial bet-hedging strategies in the presence of temporal heterogeneity and tradeoffs between multiple life-history traits and (ii) make testable predictions to stimulate future empirical work. Although we focus on the specifics of *D. discoideum*, these results might extend to the study of diversity in microbial populations in general [36]–[38], where bet-hedging has been frequently reported [39]–[42].

## 2. Model

We utilize a theoretical framework constructed by ourselves to study life-history tradeoffs in *D. discoideum* in response to environmental stressors [7] (Fig. 1A). The model is grounded in experimental observations [4]–[6] and it incorporates hypothesized life-history traits and tradeoffs and their functional consequences [43] (see Table 1 for a summary of the model assumptions). Although movement in the unicellular phase is limited and therefore cellular slime molds are likely to be found in spatially structured populations, we use a well-mixed approach in order to isolate the effects of seasonality from the effects induced by spatial structure.

In short, the model consists of growth periods, during which strain populations grow by consuming common resources, followed by starvation periods (initiated by the exhaustion of the resources), during which both spores and solitary vegetative cells die, albeit at different rates. At the transition between growth and starvation periods a population partitioning between spores and vegetative cells takes place. The starvation period ends with the arrival of a new pulse of resources, which initiates a new growth phase. Then, vegetative cells start dividing immediately, while spores have a delayed germination. This sequence of growth—starvation cycles (Fig. 1A) is repeated indefinitely, although the average length of the starvation periods will differ depending on the season.

## 2.1 Life history traits.

Following experimental observations and theoretical investigations, four traits appear to play an important role: (i) the partition between aggregated and non-aggregated cells in response to starvation [5], [6] , (ii) the number of spores, (iii) the spore size and (iv) the spore viability [4]. The first trait reflects differences among genotypes in the allocation of resources to sporulation (resulting from aggregation) or to remaining vegetative (solitary cells) and it introduces a tradeoff between dormant spores that are resistant to longer starvation periods but commit a delayed germination, and solitary cells that are less resistant to starvation but able to instantaneously start dividing upon resources replenishment. Based on experimental observations of the existence and viability of non-aggregators, this population partitioning has been theoretically hypothesized to be a bet-hedging strategy in response to fluctuations in the length of the starvation times [5], [6]. Traits (ii) and (iii), number of spores and spore size, have been experimentally shown to be negatively correlated so that a higher spore production appears to result in smaller spores [4]. Traits (iii) and (iv), spore size and viability, have been experimentally shown to be positively correlated: larger spores also show a larger viability upon germination [4]. Furthermore, spore size is inherently related to cell size, which determines cell survival [44]. These results align with work in other species in which egg or seed size determines the germination likelihood [45]–[48]. Finally, the correlations between spore size and number of spores and spore size and viability, point to a negative correlation between the number of spores and spore viability, so that a higher production of spores comes at the cost of reduced germination success for each spore.

We include these experimental observations in our model as we have done previously [7] by hypothesizing that two traits are under selection: the aggregator to non-aggregator ratio (Fig. 1B) and the division rate (Fig. 1C). The former choice follows directly from the discussion above; the latter is based on our hypothesis that differences in spore size, spore viability and number of spores, as well as correlations among them, can be entirely encapsulated by differences in the division rate among genotypes as follows: first, in the case of *D. discoideum*, spore production is directly linked to the division rate if there are no interactions in the development of the fruiting body (all strains contribute to the stalk and the spores in an equal ratio); then, negative correlations between spore viability and spore production translate into negative correlations between spore viability and division rate (Fig. 1C). Second, spore size, and implicitly cell size, can be related to the division rate via a classical growth-reproduction tradeoff between genotypes that produce a larger population made of smaller cells and genotypes that, given the same resources, produce fewer but bigger siblings (Fig. 1C). Although this is in many ways a simplifying assumption about the metabolic relationships that govern *D. discoideum*, it is consistent with classical growth-reproduction tradeoffs [43], [49] and therefore a straightforward first assumption in the absence of experimental data in slime molds and given the diverse range of relationships found in organisms across different scales [50]–[53]. Finally, given the relationship between size and viability/survival, this hypothesized growth-reproduction tradeoff results in an additional reproduction-survival tradeoff between cells that divide and consume the shared pool of resources faster but pay the cost of a decreased survival (for solitary cells) or viability (for spores), and cells that divide more slowly but have an increased survival or viability (Fig. 1C).

A condensed summary including both empirical results and the hypotheses used in the model to incorporate these experimental observations is presented in Table 1.

## 2.2. The environment

For simplicity, we will keep the initial size of the food pulses, $R_0$, and the half saturation constant, $R_{1/2}$, constant. We focus on starvation as the only environmental stressor and encapsulate the entire ecological context in the distribution of times between the onset of starvation and the arrival of the next food pulse, which we call starvation times. Alternatively, the environment could be defined in terms of the times between the arrival of food pulses. However, our results are both qualitatively and quantitatively independent of this choice (see Figure 2). This is because the time to resource depletion is always much shorter than the starvation time; thus, the distribution of starvation times remains the same regardless of whether our random variable is the starvation time itself or the time between food pulses. This is evident for large average starvation times; it is more interesting to explain why it holds for short average starvation times. Average short starvation times result in little cell death, which leads to an increase in the total population size; larger populations deplete resources faster and thus the duration of the growth period (the time to resource depletion) decreases as the mean starvation time decreases. Since these approaches lead to equivalent results, we henceforth choose the starvation time as the relevant random variable for simplicity, because it allows us to manipulate the environmental conditions independent of the growth phase.

Worse environmental conditions are represented by longer starvation times and better conditions are represented by shorter starvation times. Therefore, the environmental quality decreases with increasing starvation times. Previous studies that concentrate all ecological information into the mean starvation time use either exponential [5], [7] or uniform distributions with a fixed width [6]. Although we also study the case of exponentially distributed starvation times for comparison with previous results, we address other scenarios as well, in which the mean starvation time and the amplitude of the fluctuations can be manipulated independently. By doing this, we add an additional layer of information to the description of the environment, which is now determined not only by its average quality (as in the case of exponential or fixed-width uniform distributions), but also by how much such quality varies around its mean value, termed here environmental variability.

The uniform and the normal distribution are the most widely used distributions in which the mean value and the variance are independent. However, the necessary truncation in the distribution so that only positive starvation times are sorted, introduces a constraint in the combination of mean values and variances that can be explored using a uniform distribution. Since the truncated normal distribution does not have this limitation, we will take the starvation times to be distributed according to truncated normal distributions (see

Appendix A for a derivation of the truncated distribution starting from a standard one and Appendix B for its numerical sorting).

We consider the simplest seasonal scenario: a combination of a wet and a dry season included in the model through a periodic switch between two fixed mean starvation times indicating the dry-wet season cycle. The time series of mean starvation times shows two characteristic scales: (i) the mean starvation time within each season and (ii) the length of each season, whose inverse gives the transition rate between mean starvation times. Due to the stochasticity in the starvation times, however, the exact length of each season cannot be controlled, so we introduce two parameters, $T_{wet}$ and $T_{dry}$, that provide a lower bound to the length of the seasons rather than their exact duration. In the implementation of the model, sequences of growth-starvation cycles occur within a season until its length is exceeded. The horizontal bars in Fig. 3 provide an example of this. Seasons usually exceed $T_{dry}$ and $T_{wet}$ so the switch between them takes place at the end of the first starvation period after these values have been surpassed. In the following, however, we will refer to $T_{wet}$ and $T_{dry}$ as the length of each of the seasons for simplicity. We will study the outcome of two possible scenarios: (i) a severe dry season that consists of a very long starvation period without nutrient replenishment (Fig. 3A) and (ii) a milder dry season during which resources do replenish but are followed by starvation times that are, on average, longer than those of the wet season ($\lambda_{dry} > \lambda_{wet}$) (Fig. 3B).

## 2.3. Mathematical formulation

Each strain is defined by a pair of traits ($\alpha$, $c$): $\alpha$ represents the fraction of cells that aggregate upon starvation and $c$ represents the genotype division rate. We assume that a stochastic switch underlies the aggregator versus non-aggregator partition, such that, upon starvation, the population splits instantaneously into fixed fractions of aggregators and non-aggregators [5], [7], [44] (Fig. 1B). The limits $\alpha=0$ and $\alpha=1$ capture the extremes where either all cells remain vegetative, or they all aggregate, respectively; intermediate values represent a continuum of bet-hedging strategies. A summary of all the parameters and traits included in the model as well as their numerical values is provided in Table 2.

We assume that different strains compete for a series of pulses of resources that arrive in the environment interspersed with periods of starvation. During starvation, only cellular death takes place. The dynamics of the model starts with a growth phase during which resources, $R$, are consumed according to Monod-like dynamics [54]:

$$\dot{X}_{\alpha,c} = \frac{cR}{R + R_{\frac{1}{2}}} X_{\alpha,c} \qquad \dot{R} = \frac{-R}{R + R_{\frac{1}{2}}} \sum_{\alpha,c} Qc X_{\alpha,c} \qquad (1)$$

where $X_{\alpha,c}$ is the population size of each strain and $R_{1/2}$ is the abundance of resources at which the growth rate is half of its maximum $c$ (also termed here division rate). $Q = c_{max}/c$ is a non-dimensional conversion factor that indicates how the resources needed to produce a sibling cell scale with division rate, and implicitly with cell size [55]. Since $c_{max}$ is the division rate of the fastest dividing strain, $Q$ equals 1 for the fastest dividing strain (smallest cells) and it increases as the division rate, $c$, decreases (cell size increases) (Fig. 1C). Therefore, $Q$ defines a growth-reproduction tradeoff according to which cells consume resources at the same rate regardless of their division rate, but strains with lower division rates (larger cells) wait longer between cell divisions and hence consume more resources per division event. As non-aggregated cells only die due to starvation, we neglect the death term during the feeding phase. Once resources are exhausted, the starvation phase starts. Its duration, i.e., the starvation time $T_{st}$, is defined as the time between the exhaustion of the resources (onset of starvation) and the arrival of the next pulse of nutrients. Since following Eq. (1) resources only tend asymptotically to zero we set a threshold in the resources, $R^*$, to define the end of the growth phase. When this threshold is crossed, the starvation phase starts. Aggregation takes place instantaneously at the onset of starvation, so population splits into a fraction $\alpha$ of aggregators and a fraction 1-$\alpha$ of non-aggregators (Fig. 1B). Following experimental results that find a constant 20%:80% stalk:spore ratio for different experimental setups, we multiply the fraction of aggregated cells by a factor $s=0.8$ to obtain the number of reproductive spores. Although stalk versus spore cell allocation is an important component of *D. discoideum* life cycle, we neglect its consequences here and maintain it constant across genotypes. This is due to the fact that the stalk:spore ratio is thought to play a role in spore dispersal while here we are considering a well-mixed scenario in a single patch.

During the starvation phase there is no reproduction and cells only die; however, spores and vegetative cells show different behaviors. Dormant spores die at very small constant rate $\delta$, while vegetative cells follow a survivorship curve with a time-dependent death rate [44]. We hypothesize that all vegetative cells (regardless of size) have the same maximum lifetime but that they differ (based on size) in the probability of reaching the maximum lifespan. This assumption is grounded in previous experimental results showing that, upon food exhaustion, vegetative cells consume their own internal resources in order to survive the first hours of starvation [56]. Here we extend these results to assume that differences between genotypes are more likely to appear in the autophagy period due to differences in the availability of intracellular resources, rather than in the maximum cell lifetime. Thus, we assume that strains with higher division rate have a lower probability of reaching the maximum lifetime because they have smaller cells, with fewer intracellular resources. To implement this assumption we choose a family of survivorship curves, in which surviving probability decays continuously with time,

$$S(t) = \frac{e^{-(\mu t)^{\beta(c)}} - e^{-(\mu T_{sur})^{\beta(c)}}}{1 - e^{-(\mu T_{sur})^{\beta(c)}}} \qquad (2)$$

where $\beta(c) = 3.1-4c$ is a linear function of the division rate that accounts for the cost of having a higher division rate (reproduction-survival tradeoff, Fig. 1C). Therefore, a genotype with division rate $c$ has a specific value $\beta(c)$ that determines the temporal evolution of its survival probability; it decays faster for fastest reproducing strains. Since higher values of $\beta$ result in survivorship curves with lower short-time slope, we choose the functional form $\beta(c)$ such that: (i) $\beta(c)>1$ for all $c$ to capture the fact that survivorship curves decay slowly during the first hours of starvation (following experimental results for a single genotype [44]); and (ii) $\beta(c)$ is a decreasing function of the growth rate, to capture the survival versus reproduction tradeoff [4], [7] such that fast decaying survivorship curves correspond to strains with higher reproduction rate. $T_{sur}$ is the maximum lifetime of a vegetative cell and $\mu$ is the rate at which the survival probability of a vegetative cell decreases.

Given the death dynamics of both populations, we can obtain the number of spores and vegetative cells at the end of the starvation phase by evaluating the following expressions

$$X_{\alpha,c}^{s}(t + T_{st}) = X_{\alpha,c}^{s}(t)e^{-\delta T_{st}} \quad (3)$$

$$X_{\alpha,c}^{v}(t + T_{st}) = X_{\alpha,c}^{v}(t)S(T_{st}) \quad (4)$$

The starvation phase ends with the arrival of a new pulse of nutrients of size $R_0$. Then, surviving vegetative cells start feeding and reproducing instantaneously according to Eq. (1). Dormant spores, however, have to follow a complete germination process during which they continue to die at rate $\delta$. Once germination is completed, which takes time $\tau$, only a fraction $v(c)$ of the spores are viable and become active cells. Since, based on experimental results, we hypothesized above a negative correlation between spore viability and reproductive rate, $v$ is a decreasing function of the reproduction rate, which we here fix to be linear for simplicity, $v=1.1-2c$ (Fig. 1C). Such growth-starvation cycles continue indefinitely.

## 3. Results and Discussion.

### 3.1. One environmental scale: the emergence of bet-hedging.

We first investigate whether environmental fluctuations (stochasticity in the starvation times, i.e., times between the onset of starvation and the appearance of the next food pulse) drive the emergence of bet-hedging strategies that are defined by intermediate values of $\alpha$. A detailed description of the computational setup is provided in subsection 5.1 of Materials and Methods.

Previous studies have explored this question using either an exponential [6] or a uniform distribution of fixed width [5]. The main focus here is on truncated normal distributions that allow us to explore scenarios in which the amplitude of the environmental variability (standard deviation of the distribution of starvation times, $\sigma$) is independent of the mean starvation time (Appendix A). However, to establish a better comparison with cases in which the amplitude of the environmental variability and the mean starvation times are coupled, we also explore the outcome of exponentially distributed starvation times. To isolate the effect of temporal disorder on the bet-hedging trait we fix the value of the division rate and the tradeoffs related to it: spore viability and vegetative cell resistance to starvation (Table 2). Subsequently, in Section 3.3, we analyze the full model.

In the absence of seasonality, the temporal component has a single characteristic scale during the year, which is given by the mean starvation time $\lambda_T$. If the environment is deterministic, starvation times are fixed and coincide with this mean value (red squares in Fig. 4A and Fig. 4B; red line in Fig. 4D shows the distribution of starvation times). In this case, consistently with previous findings [5], [6], only pure strategies are selected for: $\alpha=1$ (only spores) if $T_{st} > T^*$ and $\alpha=0$ (only vegetative cells) otherwise. The transition point, $T^*=170$ hours is of the order of the maximum lifespan of vegetative cells. In the following we will refer to environments with a mean starvation time below $T^*$ as good environments and to those above this threshold as harsh environments. If the environment is stochastic (i.e.

characterized by fluctuations in starvation times), then intermediate investments in spores, which represent bet-hedging strategies, are selected for. We introduce these fluctuations by drawing the length of each starvation period from a distribution with mean $\lambda_T$. Two classes of distributions are investigated; (i) exponential distributions (Fig. 4A, 4B) and (ii) a family of truncated normal distributions with different amplitudes (Fig. 4C, 4D). The truncation of the distribution, with a cutoff at $T_{st} = 0$ avoids unrealistic negative values for the starvation times. In addition, due to the cutoff at $T_{st} = 0$, the mode of the distribution (i.e., the most frequent values for the starvation times) decreases when the standard deviation increases and the mean value is kept constant.

Consistent with previous results [5], [6], we find that higher investments in spores are favored as the environments become harsher (i.e., characterized by larger mean starvation times) (Fig. 4A and 4C). For the truncated normal distribution, increasing the amplitude of the variation in the starvation times promotes the evolution of bet-hedging both in good and in harsh environments. In harsh environments, however, bet-hedging requires higher amplitudes. Strategies with a higher investment in nonaggregators are riskier and therefore the effect of long starvation times on them is stronger than the effect of short starvation times on strategies with higher investment in spores. Due to the truncation at $T_{st} = 0$, the mode of the distribution approaches zero as its variance increases. This makes short starvation periods more frequent, which penalizes a pure strategy with $α=1$. In fact, the exponential distribution can be seen as a limit case of a truncated normal distribution with a very high standard deviation (see orange line in Fig. 4D and the distribution in Fig. 4B), which explains the higher impact of exponential variability on the winning strategy in harsh environments (black squares in Fig. 4A).

### 3.2. Two environmental scales: the effect of seasonality on strain coexistence.

The general setup of the simulations, outlined in subsection 5.1 of Materials and Methods, is essentially the same as in the non-seasonal case studied in Section 3.1. Only the environmental conditions are changed to include seasonality according to Section 2.2. In a first exploration using exponential distribution for the starvation times in both seasons, coexistence was not found. Therefore, in the following sections we will use a truncated normal distribution for each and discuss the reasons why exponentials cannot lead to coexistence.

### 3.2.1 Severe dry season: no resource replenishment.

We first investigate the scenario in which resources are not replenished in the system during the dry season; the entire dry season is one starvation period. The wet season, however, consists of several growth-starvation cycles with starvation times sorted either from a truncated normal distribution of mean value $\lambda_{wet}$ and standard deviation $σ$ or from an exponential distribution of mean value $\lambda_{wet}$. We will focus on the effects of the dry season length and of the mean starvation time in the wet season. Therefore, the parameters of interest are $T_{dry}$ and $\lambda_{wet}$ (and $σ$ for truncated normal distributions), while $\lambda_{dry} = T_{dry}$ and $T_{wet} = 1$ year - $T_{dry}$ are determined by the length of the dry season. As explained above, the parameters $T_{dry}$ and $T_{wet}$ provide a lower limit for the length of the seasons. Given this setup, the distribution of starvation times over many years has two components: (i) a Dirac delta distribution centered at $\lambda_{dry}=T_{dry}$ for the dry season and (ii) a truncated normal distribution (respectively exponential) of mean value $\lambda_{wet}$ and standard deviation σ for the wet season (Fig. 3A).

Regardless of the distribution used, the periodic occurrence of very long starvation times pushes selection towards higher investment in spores, the higher the longer the dry season is (Fig. 5). This effect is however residual if the wet season covers at least 25% of the year (blue squares in Fig. 5). This result follows from the fact that sporadic long starvation times kill all vegetative cells and dramatically decrease the total population of spores, which favors strains with a higher value of α. When spores and non-aggregators constitute a bet-hedging strategy, the change in the optimal value of α implies that catastrophic ecological periods promote the evolution of lower-risk strategies, represented by the production of more spores. However, coexistence of bet-hedging strategies is not possible since there is no reproduction during the dry season and thus a second strategy cannot evolve. It is interesting to note that even purely deterministic scenarios in which starvation times during the wet season are constant, allow the evolution of bet-hedging strategies (Fig. 5A). The success of bet-hedging strategies even in deterministic environments arises from the alternation in the environmental conditions that reduces the fitness of pure strategies, making them suboptimal compared to a mixed investment in aggregators and non-aggregators. Finally, when the wet season is stochastic with starvation times drawn from a truncated normal distribution, the set of environments dominated by bet-hedging strategies increases with the variance of the distributions, similarly to the case without seasonality (Fig. 5B, C).

### 3.2.2 Milder dry season: slow food recovery.

Here we study a less extreme ecological scenario in which pulses of resources also arrive during the dry season, albeit with a lower frequency ($\lambda_{dry}>\lambda_{wet}$). Seasonality is now illustrated by a periodic switch between two distributions with different mean starvation times (Fig. 3B). To simplify the analysis, we fix the length of the seasons and divide the year into 6 months of wet favorable conditions and 6 months of dry harsh environments. To reduce the dimensionality of the parameter space we also fix the mean starvation time in the wet season, $\lambda_{wet} = 50$ hour, and assume that the intensity of the environmental fluctuations, $\sigma$, is the same during both seasons. Although a more realistic approach should account for differences in $\sigma$ between seasons, this assumption reduces the dimensionality of the parameter space and the complexity of the analysis without qualitatively changing the results presented in this section. The number of parameters reduces thus to two: the mean starvation time in the dry season, $\lambda_{dry}$, and the intensity of the fluctuations in both seasons, $\sigma$. We find that two strategies, a wet season and a dry season specialist, coexist if two conditions are met (Fig. 6). (i) Temporal niche partition: all starvation times (not only their mean values) within a season are similar among themselves and different to those in the other season, such that two niches are created within the year. This condition requires variance in the starvation times within each season to be low. (ii) The number of growth-starvation cycles that occur within each season is sufficiently large for its specialist strain to create a large population that is able to survive when the environment changes and the strain becomes maladapted. These conditions are violated, for instance, if exponential distributions are used for both seasons. Since $\lambda_{wet}$ is necessarily lower than $\lambda_{dry}$, if $\lambda_{dry}$ is too low, then the starvation times in both seasons are too similar and the first condition for coexistence is violated. Since mean value and standard deviation are entangled in exponential distributions, increasing $\lambda_{dry}$ also increases the variance in the season's starvation times. The mode of the distribution remains, however, at zero. Therefore, the overlap between dry and wet season distributions of starvation times remains large and the first condition is violated again. Finally, in the limit in which the dry season mean starvation time is large, very few growth-starvation cycles take place within the season and the second condition is violated. Eventually when $\lambda_{dry}$ reaches an upper bound, a single starvation time will cover the whole dry season and the extreme scenario introduced in Section 3.2.1 is recovered. However, two normal distributions or a combination of an exponential distribution in one season and a normal one in the other season can enable coexistence, even though the parameter range in which that will occur will depend on the distributions and the rest of the model parameters.

The need for temporal niche partitioning and persistence of season specialists through unfavorable periods shows that coexistence is driven by a temporal storage effect. The third requirement in [12], [25], [32] related to the covariance between interspecific competition and environmental changes, is also implicitly fulfilled: transitions in the seasons penalize the most abundant genotype, whose population starts declining; this reduces interstrain competition and allows the new season specialist to recover from a very low population size. As a result, there is a stable coexistence of two strategies that oscillate with a period that depends on the length of the seasons (Fig. 6D) [18].

Temporal niche partitioning requires an upper bound in the intensity of the fluctuations. Environments with high variability result in wider distributions for the starvation times, which may increase the similarity between the distributions of wet and dry season starvation times. This tends to eliminate the temporal niche partitioning and, in consequence, coexistence is lost. Although changes in the variance of the truncated distribution can also modify the shape of the distribution (especially for low values of the mean), this has no impact on the loss of coexistence since only the wet season distribution changes significantly by accumulating more starvation times closer to zero (see Fig. 4D for an example of how the shape of the distribution changes for large values of the variance). More frequent very short starvation times reinforce coexistence as they are more disparate to dry season starvation times. Therefore, the longer tail of the wet season distribution, caused by a larger variance, leads to a loss of coexistence by increasing the variability in the wet season starvation times.

On the other hand, for fixed environmental variability (i.e. the standard deviation of the distributions of starvation times), starvation times within each season are more similar to those within the other season as the mean values get closer. Since we keep the mean starvation time of the wet season, $\lambda_{wet}$, constant, this introduces a lower limit for the mean starvation time of the dry season in order for coexistence to be maintained. In addition, the requirement for several growth-starvation cycles per season excludes extremely high mean starvation times from the region of coexistence in the parameter space (Fig. 6A). Therefore, we expect coexistence for intermediate values of $\lambda_{dry}$.

Finally, we study how surviving strategies vary as the ecological variables (i.e. dry season mean starvation time, $\lambda_{dry}$, and amplitude of the environmental variability, $\sigma$) change. Due to the fixed length of the seasons, the number of growth-starvation cycles that they permit decreases as the mean starvation time increases. This reduces the number of generations and the evolutionary impact of a season on the winning strategy. For the case studied in Figure 5 (Fig. 6B, dashed-line transect of Fig. 6A), in which the length of both seasons is fixed to 6 months and the mean starvation time in the wet season is fixed to 50 hours, the number of generations within the dry season decreases as $\lambda_{dry}$ increases. Consequently, selection favors strategies with a lower investment in spores and coexistence is eventually lost in the limit of high $\lambda_{dry}$, in which very few reproduction events take place during the dry season. This result should not be confused with the scenario tackled in Section 3.2.1, in which higher values of $\lambda_{dry}$ lead to investment in more spores. In that case, season length is determined by $\lambda_{dry}$, with higher values of $\lambda_{dry}$ leading to larger dry seasons and shorter wet periods. Regarding to the environmental variability, medium to higher values of σ force the loss of one of the strains (Fig. 6C, dashed-line transect of Fig. 6A). Higher dispersion in the wet and dry season starvation times (more overlap between the starvation times distributions), increases the similarity of starvation times between seasons and two well differentiated niches within the year are lost. Consequently, one of the evolving strategies has a sufficiently high fitness in both seasons to outcompete specialist strategies. The winner always evolves from the former wet season specialist because most of the growth-starvation cycles occur in that part of the year.

## 3.3. Diversity promoted by additional life-history tradeoffs.

Finally, we consider the full model, in which the division rate $c$ is an evolving trait that contributes to the fitness of each genotype and establishes to two additional tradeoffs [4]: one between spore viability and division rate and another between starving solitary cell survival and division rate, both mediated by an underlying growth-reproduction tradeoff between division rate and cell size. In other words, genotypes with higher division rates produce smaller cells that are less resistant to starvation and smaller spores that are less viable upon germination (Fig. 1B). Since division rate is now a trait that can evolve, the setup of the simulations is modified to include a larger set of strains (see subsection 5.2 in Materials and methods for details).

In this complete framework, we find that a third strategy can coexist with the two season-specialist bet-hedging strategies. The third strain is an annual generalist strategy, called henceforth an annual bet-hedger, whose division rate is lower than that of the season-specialists. The annual bet-hedger first grows at the beginning of each season but soon starts declining due to interspecific competition with the season specialist. This occurs when the dry season has a low mean starvation time and both seasons have low variability (Fig. 7A-E), since the persistence of an annual generalist strategy requires seasons to be sufficiently similar (though not too similar to avoid loss of specialist coexistence), and to have a low level of environmental variability. At higher environmental variabilities, differences between seasons are attenuated by the effect of the fluctuations; then, the specialists are able to survive at higher population sizes during adverse periods and they outcompete any annual bet-hedger. Eventually, as we already showed in Section 3.2.2, as the intensity of the environmental fluctuations keeps increasing, temporal niche partitioning is lost and a single strategy emerges and dominates throughout the year.

In this full model, genotypes behave differently both in the growth and in the starvation phase and the length of the growing periods, in addition to the number of growth-starvation cycles and the length of the starvation phases, plays an important role in determining the outcome of the competition. All three variables determine the scenarios in which dividing faster (or slower) is beneficial, i.e., when it is better to produce a larger population albeit consisting of less resistant vegetative cells and less viable spores, and when it is better to invest in a smaller population consisting of more resistant cells and more viable spores. The duration of the growth period, i.e., the time between the arrival of the food pulse and its exhaustion, is determined by the size and the strain composition of the total population and it can be used as an estimator for the strength of the competition for resources: the smaller the total population size, the lower the competition for resources and the larger the exponential growth phase. Therefore, scenarios in which the total population is very small favor strains with fast division rates, since they can produce a sufficiently large population during the growth phase and compensate the costs associated with lower cell resistance and spore viability. Even though more cells with a larger $c$ will die during the starvation phase and a smaller fraction of spores will germinate upon resource replenishment, these strains will still be overrepresented in the population because they produced a much larger population during the growth phase and they could balance these losses. However, as the total population size increases, the growth phases become shorter and the strains with a higher division rate cannot create such a large population. In those cases, selection favors genotypes that reproduce more slowly but whose offspring have an increased survival/viability.

Following the previous rationale, we can explain the behavior of the optimal division rate in Fig. 7C and 7E. At the beginning of the dry seasons, the total population size is large due to the favorable conditions provided by the previous wet season, so the annual bet-hedger, which has the smallest division rate, is favored [7]. Then, as the wet-season specialist declines, the feeding periods become longer and the dry-season specialist, with a higher division rate, starts outcompeting the annual bet-hedger. On the other hand, at the beginning of the wet season, the population of the wet season specialist is low because it has declined during the previous dry season and its vegetative cells do not consume most of the resources before spores germinate. As a consequence, the dry-season specialist and the annual generalist do not pay a significant cost for their high value of $\alpha$ since the majority of resources are still in system when their spores germinate and their populations can still grow. However, since the annual bet-hedger has lower division rate and therefore a higher spore viability, it outcompetes the dry-season specialist, whose population declines. Finally, as the wet season progresses, the wet season specialist grows, a higher investment in spores is penalized and the annual bet-hedger declines as well.

Since the annual bet-hedger grows and declines during each season its population oscillates with a frequency that is approximately twice the frequency of the oscillations in the season-specialist populations (inset of Fig. 7F). Finally, the optimal investment in spores (Figure 7B and 7D) behaves as in the simpler case where division rate is fixed, and it can be explained using the same arguments introduced in Section 3.2.2.

# 4. Conclusions.

We investigated theoretically the scenarios under which temporal heterogeneity, here captured through the existence of a wet and dry season, drives the coexistence of competing bet-hedging strategies in microbial populations. We further explored the interplay between coexistence and the number of life history traits and the tradeoffs among them. Our main focus was on the cellular slime mold *D. discoideum*, which has recently been proposed as an example of bet-hedging in microbes [5], [6], [39] and shows a vast biodiversity [3]. However, the model can be easily extended to other microbes and the results are likely general.

The evolution of bet-hedging strategies is driven by environmental stochasticity for various sources of fluctuations, such as exponential [5], uniform [6] and, as studied here, truncated normal distributions. For the latter, the effect of the mean value and the amplitude of the fluctuations can be teased apart to show that more intense fluctuations select for strategies with a higher investment in spores. In this paper, we included seasonality as an additional temporal scale on environmental variability to determine what are the minimal necessary conditions under which a single strategy is not able to average over the different environmental contexts and is replaced, via temporal niche partitioning and a temporal storage effect, by coexisting season specialists.

We investigated two different patterns of seasonality determined by the properties of the dry season. First, a severe dry season represented by a single starvation period that spans the entire season. In this scenario, since reproduction only takes place in the presence of food, wet seasons permit many more generations than dry seasons, which only act as catastrophic perturbations that favor selection for higher investment in spores (lower-risk strategies). This result is consistent with previous studies in annual plants that showed how catastrophic years favor the evolution of lower risk germination strategies [27]. Second, we studied a milder dry season that allows for several growth cycles, albeit fewer than the wet one. We showed that this can drive the evolution of season-specialist bet-hedging strategies that stably coexist via a temporal storage effect [12], [25], [32]. In our model, two conditions must be fulfilled: (i) all the starvation times within a season have to be similar among themselves and different to those in the other season in order to create two different habitats within the year (temporal niche partition), and (ii) each season has to accommodate numerous growth-starvation cycles in order to create a storage of each strain and guarantee its persistence through adverse conditions.

An extended model that considers additional genotypic traits under selection and multiple tradeoffs mediated by these traits enables the coexistence of more than two strains. This is consistent with existing results that emphasize the importance of tradeoffs in establishing species coexistence [12], [34], [37], [57], [58]. The additional trait we considered was the division rate, which establishes a growth-reproduction tradeoff such that strains that divide faster produce more but smaller cells. On the other hand, a slow division rate translates into the production of fewer but bigger sibling cells. Such a tradeoff mediates additional tradeoffs between division rate and spore viability – smaller spores are less viable – and between reproduction and survival – bigger vegetative cells with lower division rates are also more resistant to starvation [4]. In addition to the dry and the wet season specialists, a third annual bet-hedging strain is able to persist by growing at the beginning of each season, when interspecific competition is low and that the number of growth-starvation cycles within each season prevents either of the specialists from completely outcompeting it by the end of the season. This is consistent with previous studies in annual plants, where a tradeoff between seed survivorship and seed yield was shown to increase the set of environments in which coexistence is possible [27]. However, we extended these existing results and show that not only the number of environments can increase, but also the number of coexisting species. Our results highlight the importance of ecological variables along with multiple fitness components and tradeoffs among them in explaining long-term strain coexistence in *D. discoideum*.

Several future directions expand upon this study. First, the well-mixed approach presented here neglects the effect of spatial degrees of freedom that may also play an important role [5], [59]–[63]. Exploring whether a combination of temporal and spatial heterogeneity expands the set of ecological scenarios in which coexistence may be found as well as the number of coexisting species arises as a natural extension of this work. Second, our framework omits the existence of cell-to-cell interactions that occur during aggregation and fruiting body development; including these could modify the investment in spores in mixed populations as compared to clonal scenarios [64] and should be investigated in future work. Third, our model does not include mutation, meiotic recombination or horizontal gene transfer, all of which could reinforce strain coexistence. Although mutation rates are very low in *D. discoideum* [8], meiotic recombination rate seems to be sufficiently large to influence population composition [65]. Fourth, an important direction for future research should focus on the exploration of different tradeoff implementing mechanisms, aiming to generalize the model to other microbes. Particularly interesting is the effect of cell size on the maximum resources uptake rate and division rate, here accounted via a simple growth-reproduction tradeoff. Although a large literature has found power-like relationships between resources intake rate and cell size, recent studies in phytoplankton have reported the existence of a tradeoff between nutrient uptake and metabolism that is reflected by non-monotonic relationships between maximum growth rate and cell volume [50]. The general mechanisms that regulate the relationship between cell size and division rate across different scales and especially in microbes are thus unclear [53], [66] and have not been tested in *D. discoideum*. Finally, diversifying the sources of ecological uncertainty between the starving (mean starvation times) and the growing phase (amount and quality of the resources together with strain specific dietary preferences [67] ) would provide a more complete picture of the ecological forces that drive the coexistence of bet-hedging strategies.

## 5. Materials and Methods

### 5.1. Numerical simulations with fixed division rate

We initialize the system with 1001 different strains given by their aggregator to non-aggregator ratio, *α,* and run numerical simulations of the growth-aggregation-differentiation-starvation sequence defined by the model until time exceeds $2 \times 10^8$ hours. Strains are taken so that the values of *α* are uniformly distributed in the interval [0,1]. The initial abundance of each strain is independently drawn from a standard log-normal distribution and subsequently normalized so that the entire population contains $10^8$ cells.

At the end of the integration time, we measure the genotype abundance distribution and determine the winning genotype as the most abundant one. For computational feasibility, we stop the simulation at $t=2\times10^8$ hour, when, although one genotype is much more abundant than the others, several strains still persist in the system. However, numerical simulations over longer times show that, in the absence of seasonality, only one strain survives if simulations are run for a sufficient time. When seasonality is included, to evaluate the coexistence of several genotypes and their mean frequency in the population, we average the abundance of each strain at the end of every season from $t=1.5\times10^8$ hour to $t=2\times10^8$ hour normalized by the total population. When two strains coexist, the distribution in the strain frequency becomes bimodal, reflecting the coexistence of two genotypes. In addition to temporal averages, when the starvation times are stochastic, mean values are also obtained over 20 independent realizations. A single realization is used for deterministic starvation times.

## 5.2. Numerical simulations of the complete model

All the simulations start with an initial population compounded by 401401 strains (1001 values of $\alpha$ between 0 and 1 and 401 values of $c$ between 0.05 and 0.45). The initial abundance of each genotypes is also drawn from a log-normal distribution and subsequently normalized so that the total population size is equal to $10^8$ cells. Coexistence and the frequency of each strain in the population is evaluated with the protocol followed for the model with fixed division rate.

## Acknowledgments

This work has been funded by the Gordon & Betty Moore Foundation through grant GBMF2550.06 to RMG and the Sloan Foundation through grant FR-2015-65382 to CET. We thank the IFISC (CSIC-UIB) computing lab for technical support and the use of their computational resources. We are grateful to Joan E. Strassmann for suggesting the study of seasonality in *D. discoideum* and to George W.A. Constable and Juan A. Bonachela for useful discussions.

## Appendix A. Generation of truncated normal distributions

A truncated normal distribution is usually defined in two steps: (i) choosing a standard normal distribution, called parent distribution, of parameters $(\bar{\mu}, \bar{\sigma})$ and (ii) specifying a truncation range $(a, b)$. The probability density function (PDF) of the truncated distribution is obtained by setting the values of the original PDF to zero outside the truncation range and uniformly rescaling the values inside the range so that the norm is 1. The truncation changes the mean value and the standard deviation of the original normal distribution. These new values correspond in the model with the mean starvation time, $\lambda_T$, and the environmental variability amplitude, $\sigma$, respectively.

The truncated distribution will be symbolized by $\psi(\bar{\mu}, \bar{\sigma}, a, b; t)$ and its parental standard normal distribution by $N(\bar{\mu}, \bar{\sigma}; t)$. We will work with $a = 0$, $b = 1$ and $t$ will be the length of the starvation period. To obtain which distribution has the appropriate mean value, $\lambda_T$, and standard deviation, $\sigma$, the parental distribution has to be obtained solving:

$$\int_0^\infty \psi(\bar{\mu}, \bar{\sigma}, a, b; t)\, dt = \int_0^\infty K N(\bar{\mu}, \bar{\sigma}; t) dt = 1 \tag{A.1}$$

$$\int_0^\infty \psi(\bar{\mu}, \bar{\sigma}, a, b; t)\, t dt = \int_0^\infty K N(\bar{\mu}, \bar{\sigma}; t) t dt = \lambda_T \tag{A.2}$$

$$\int_0^\infty \psi(\bar{\mu}, \bar{\sigma}, a, b; t)\, t^2 dt - \lambda_T^2 = \int_0^\infty K N(\bar{\mu}, \bar{\sigma}; t) t^2 dt - \lambda_T^2 = \sigma^2 \tag{A.3}$$

that gives the following system of equations that can be solved for each pair of $(\lambda_T, \sigma)$ values

$$\frac{1}{2} K \left[ 1 + \mathrm{Err}\left(\frac{\bar{\mu}}{\sqrt{2}\bar{\sigma}}\right) \right] = 1 \tag{A.4}$$

$$\frac{1}{2} K \left[ \bar{\mu} + \frac{\sqrt{2}}{\pi} \bar{\sigma} \exp\left(-\frac{\bar{\mu}^2}{\sqrt{2\bar{\sigma}^2}}\right) + \bar{\mu} \mathrm{Err}\left(\frac{\bar{\mu}}{\sqrt{2}\bar{\sigma}}\right) \right] = \lambda_T \tag{A.5}$$

$$\frac{\bar{\mu}\bar{\sigma}}{\sqrt{2\pi}} K \exp\left(-\frac{\bar{\mu}^2}{\sqrt{2\bar{\sigma}^2}}\right) + \frac{1}{2} K(\bar{\mu}^2 + \bar{\sigma}^2) \left[ 1 + \mathrm{Err}\left(\frac{\bar{\mu}}{\sqrt{2}\bar{\sigma}}\right) \right] - \lambda_T^2 = \sigma^2 \tag{A.6}$$

where $\mathrm{Err}(x) = \sqrt{\frac{2}{\pi}} \int_0^x e^{-t^2} dt$ is the error function. To solve numerically Eqs.(A.5) and (A.6) we use an approximation to the error function given by [68]:

$$\mathrm{Err}(x) = \begin{cases} -1 + \sum_{i=1}^{5} a_i f(x)^i e^{-x^2} & \text{if } x \leq 0 \\ 1 - \sum_{i=1}^{5} a_i f(x)^i e^{-x^2} & \text{if } x > 0 \end{cases} \tag{A.7}$$

where $f(x) = \frac{1}{(1+p|x|)}$ with $p = 0.32759$. The values of the coefficients are $a_1=0.254829$, $a_2=-0.284496$, $a_3=1.421413$, $a_4=-1.453152$, $a_5=1.061405$.

## Appendix B. Simulation of the truncated normal distribution

We use two methods to simulate the truncated distribution:

    1. If $\bar{\mu} \geq 0$, we sort random numbers following the parental normal distribution. They are rejected whenever they are negative and a new number is sorted until getting a positive value.

    2. If $\bar{\mu} \geq 0$, the previous method may be extremely inefficient as the number of rejections increases with decreasing $\bar{\mu}$. In this case we use an acceptance-rejection method [69]. The basic idea is to find an alternative probability distribution $G$ such that we already have an efficient algorithm for generating random numbers and that its density function $g(t)$ is similar to the truncated normal distribution, $\psi(\bar{\mu}, \bar{\sigma}, a, b; t)$. The steps of the algorithm are:

    (a) Generate a random number $U_1$ distributed as $G$.

    (b) Generate a random number $U_2$ from a uniform distribution between [0, 1].

    (c) If:

$$U_2 \leq \frac{\psi(U_1)}{g(U_2)} \tag{B.1}$$

    then accept $U_1$; otherwise go back to (a) and generate a new number.

    Therefore, in order to minimize the number of rejections, the ratio $\psi(\bar{\mu}, \bar{\sigma}, a, b; t)/g(t)$ has to be lower but as close as possible to 1 for all $t$. We work with $g(t)=1.3N(\bar{\mu}, \bar{\sigma};0)\exp(-0.01t)$, which approximates the tail of the parental normal distributions $N(\bar{\mu}, \bar{\sigma};t)$ when $\bar{\mu}<<0$ and gives a high percentage of acceptances.

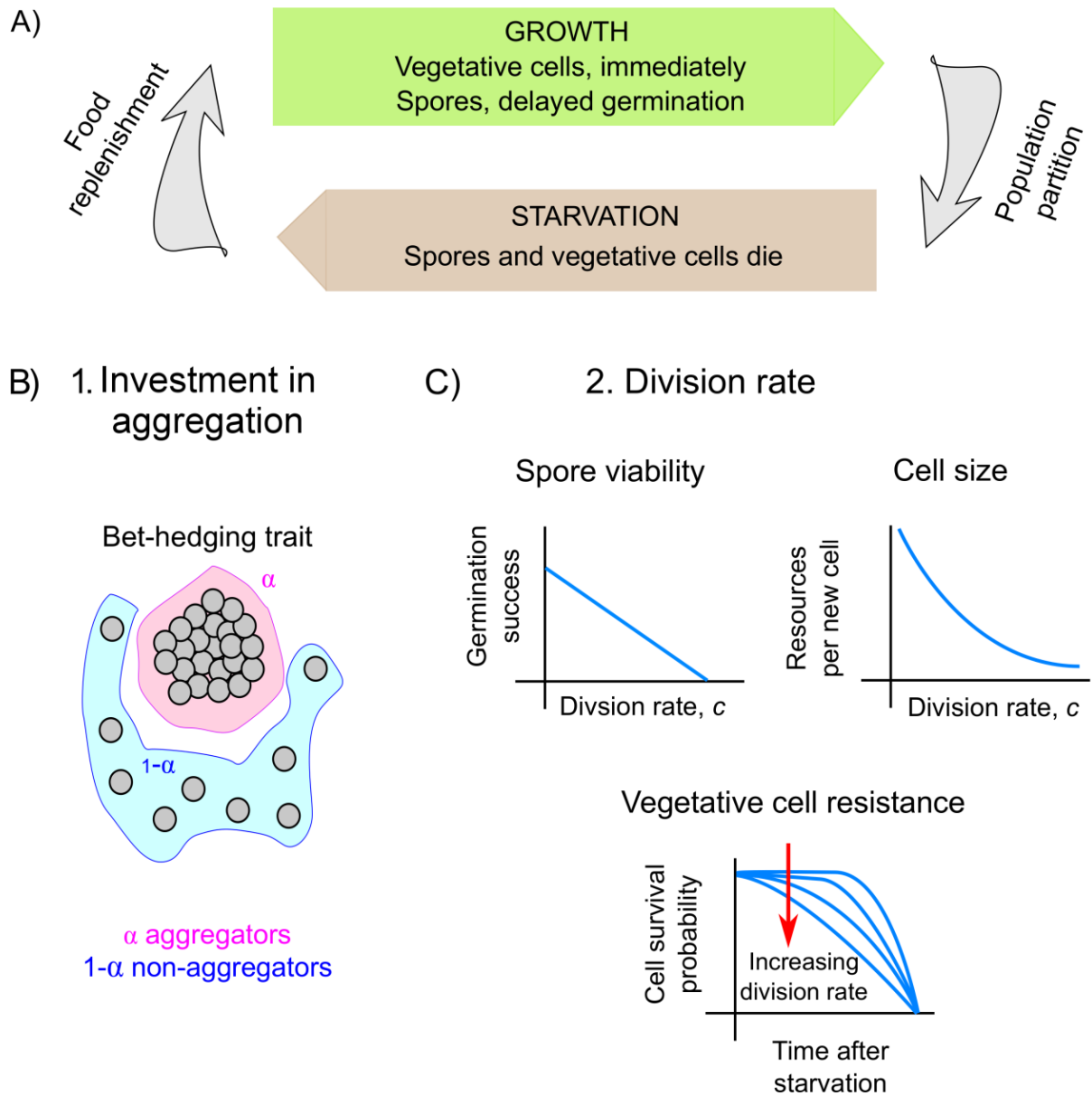

**Fig 1.** *D. discoideum* life cycle and fitness tradeoffs. A) The life cycle of *D. discoideum* consists of a series of growth and starvation periods. The growth phase starts with the arrival of a pulse of resources and aggregation, which in our model is represented by a population partition between spores and vegetative cells followed by spore:stalk cell differentiation, occurs immediately after resources exhaustion. B) A tradeoff between the commitment to aggregating or staying vegetative determines the bet-hedging strategy of each genotype. C) Three additional tradeoffs yield between division rate and spore viability (reproduction-survival tradeoff), between division rate and resistance to starvation (reproduction-survival tradeoff), and between division rate and cell size (growth-reproduction tradeoff).

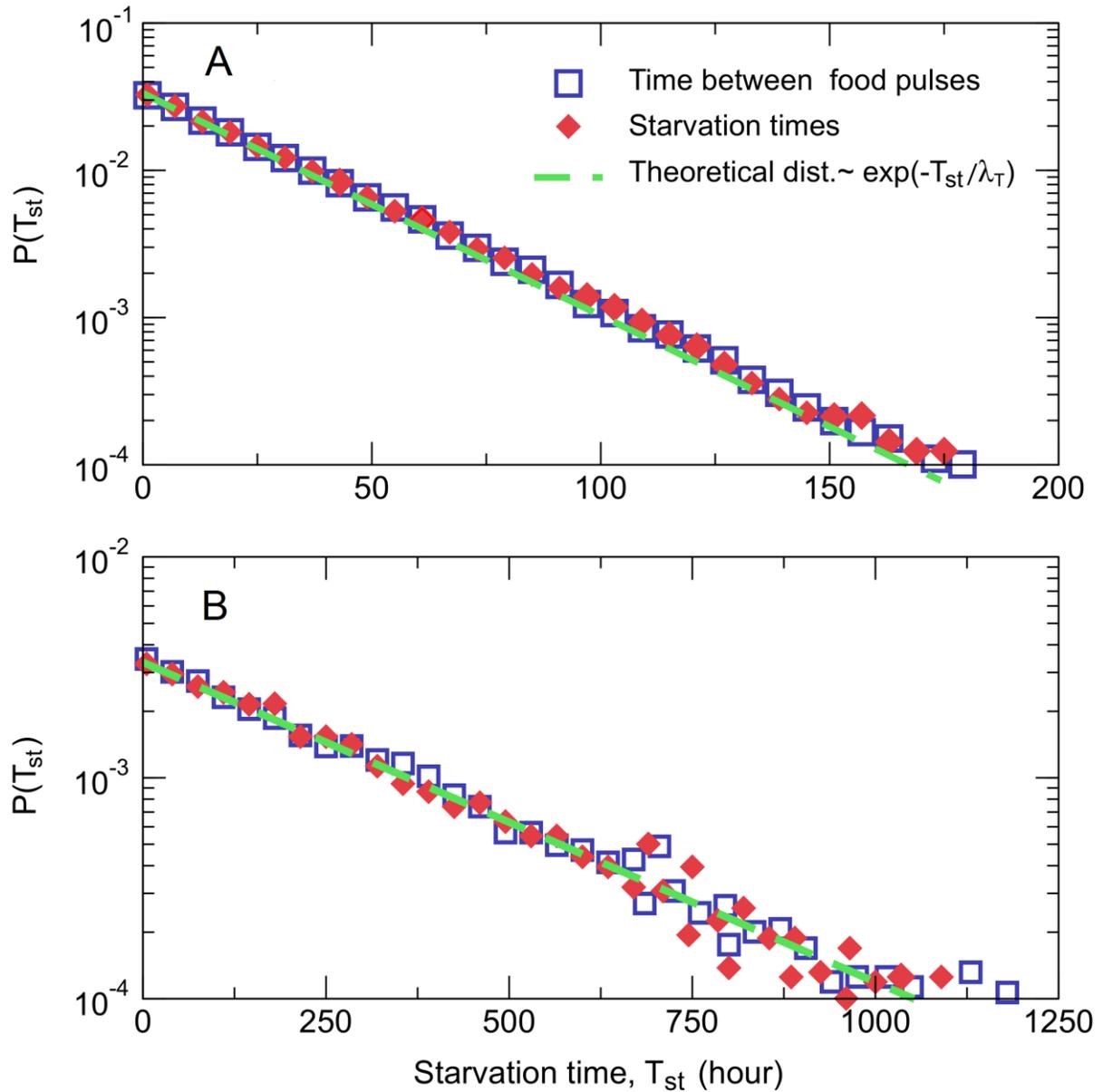

**Fig 2.** Distribution of the length of the starvation periods obtained from numerical simulations in which the starvation time (red diamonds) or the time between food pulses (blue empty squares) is the random variable. The green lines show the theoretical exponential distribution of mean value (A) $\lambda_T=30$ hours and and (B) $\lambda_T=300$ hours (low panel).

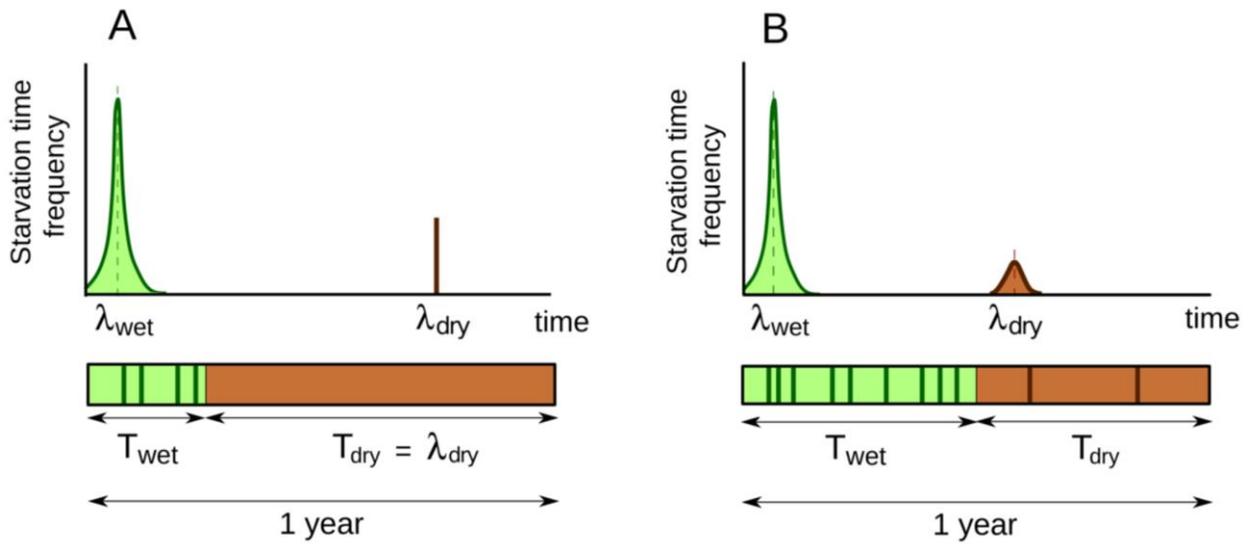

**Fig 3.** Implementation of the wet-dry annual seasonality. Symbols: $T_{wet(dry)}$, length of the wet(dry) season; $\lambda_{wet(dry)}$, mean starvation time in the wet (dry) season. A year is represented by the lower bar that is divided into a dry and a wet season by black vertical lines. Food pulses are represented by a vertical (brown or green) solid lines. A) Severe starvation. The length of the dry season varies to explore its effect on the winning genotype. During the dry season, represented by the brown segment, resources do not recover in the limit of $\lambda_{dry}=T_{dry}$. In the wet season, however, starvation times follow a truncated normal distribution (respectively exponential, not shown) of mean $\lambda_{wet}$. This combination results in a normal distribution for the wet season starvation times and a Dirac delta distribution for the dry season starvation times. B) Milder dry season that enables nutrient replenishment. The length of both seasons is fixed to 6 months and pulses of resources arrive to the patch with different frequencies in each sesson, $\lambda_{wet} < \lambda_{dry}$.
27

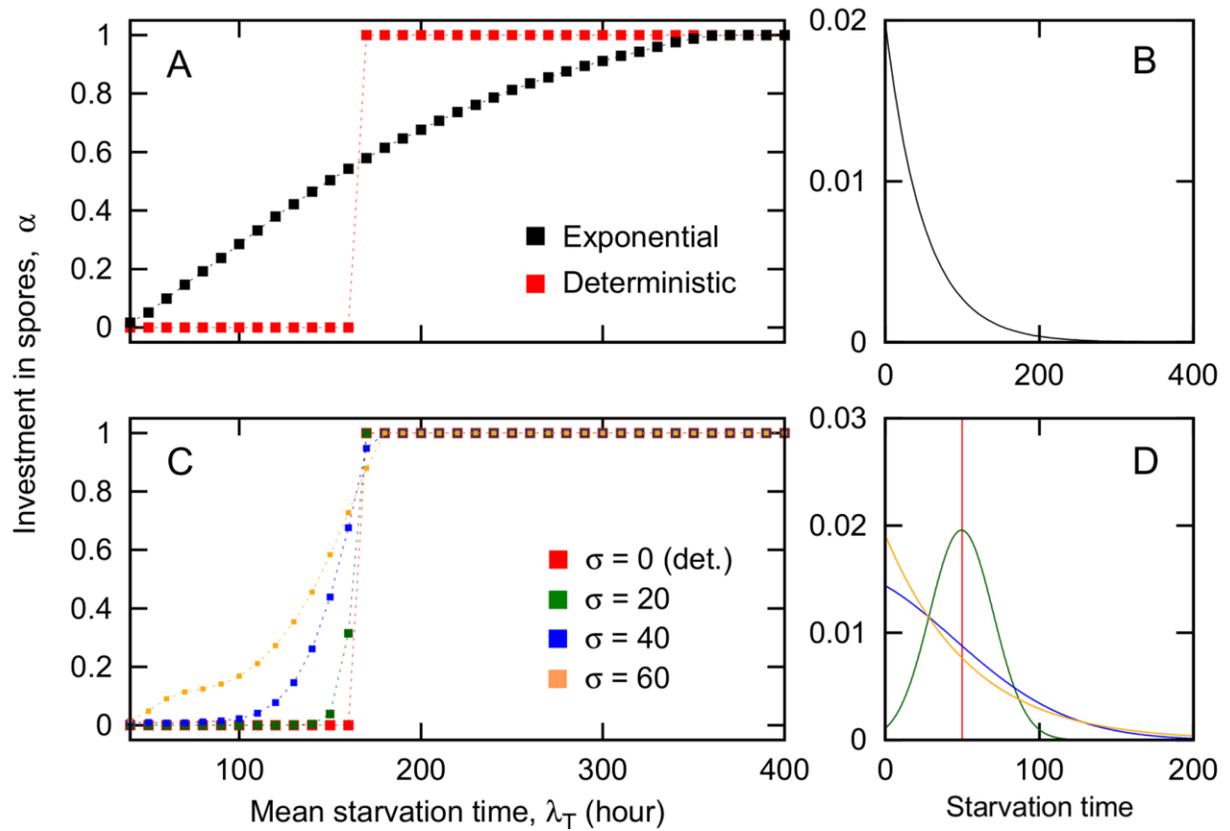

**Fig 4.** Winning genotype in deterministic and stochastic environments. The squares in panels A and C represent the optimal investment in spores and the dashed lines are interpolations. Simulations are started with 1001 genotypes uniformly distributed between $α=0$ and $α=1$ and winners are obtained as explained in Materials and Methods. A) Red squares correspond to the deterministic case and black squares to exponentially distributed starvation times. B) Starvation times exponential distribution of mean value $λ_T=50$ hour. C) From top to bottom (increasing square size): $σ=60$ (orange), $σ=40$ (blue), $σ=20$ (green), $σ=0$ deterministic (red). D) Starvation times truncated normal distributions of mean value $λ_T=50$ hour and increasing σ (same color code as in panel C). The red line represents a Dirac delta distribution.

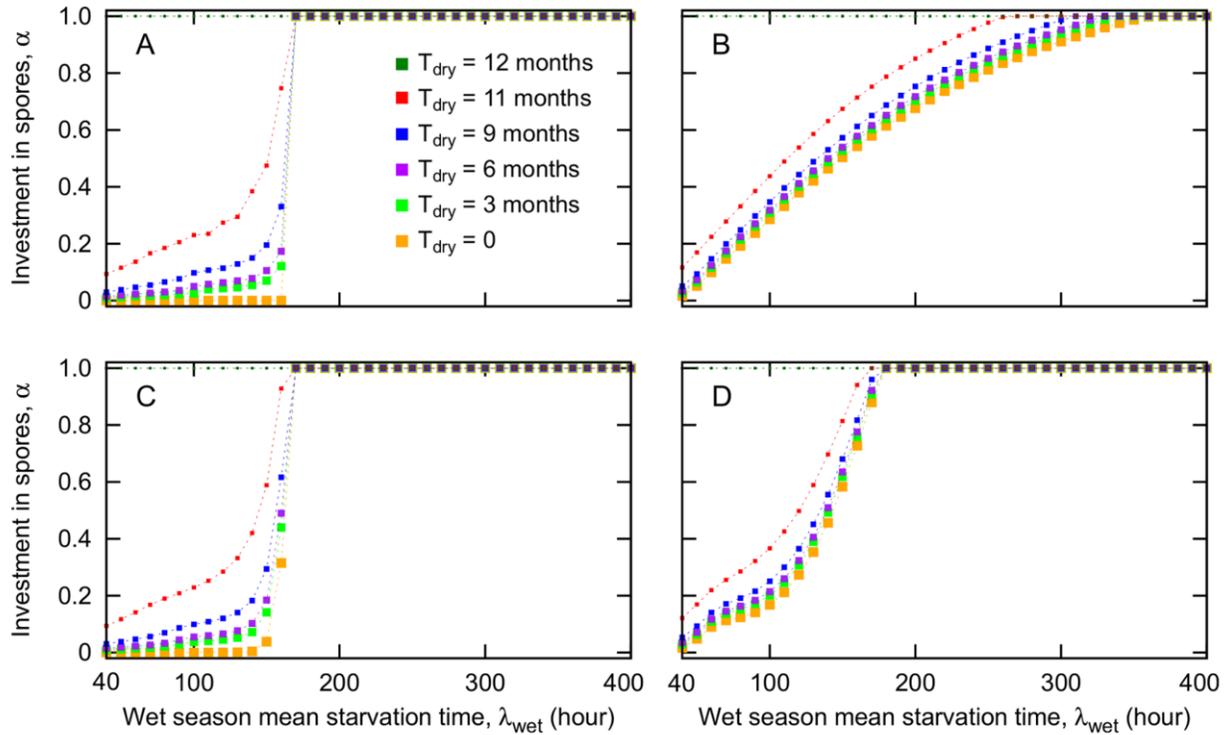

**Fig 5.** Effect of a severe dry season ($T_{dry}= \lambda_{dry}$) on the winning genotype. A) Deterministic wet season. B, C) Environmental variability follows a truncated normal distribution, with $\sigma=20$ (B) and $\sigma=60$ (C). D) Exponential distribution for the environmental quality. Simulations are initialized with 1001 genotypes uniformly distributed between $\alpha=0$ and $\alpha=1$. Distinct colors correspond to different lengths of the dry season, from bottom to top: no dry season (orange), 3 months (green), 6 months (purple), 9 months (blue), 11 months (red) and no wet season (dark green). Winners are obtained as explained in Materials and Methods.

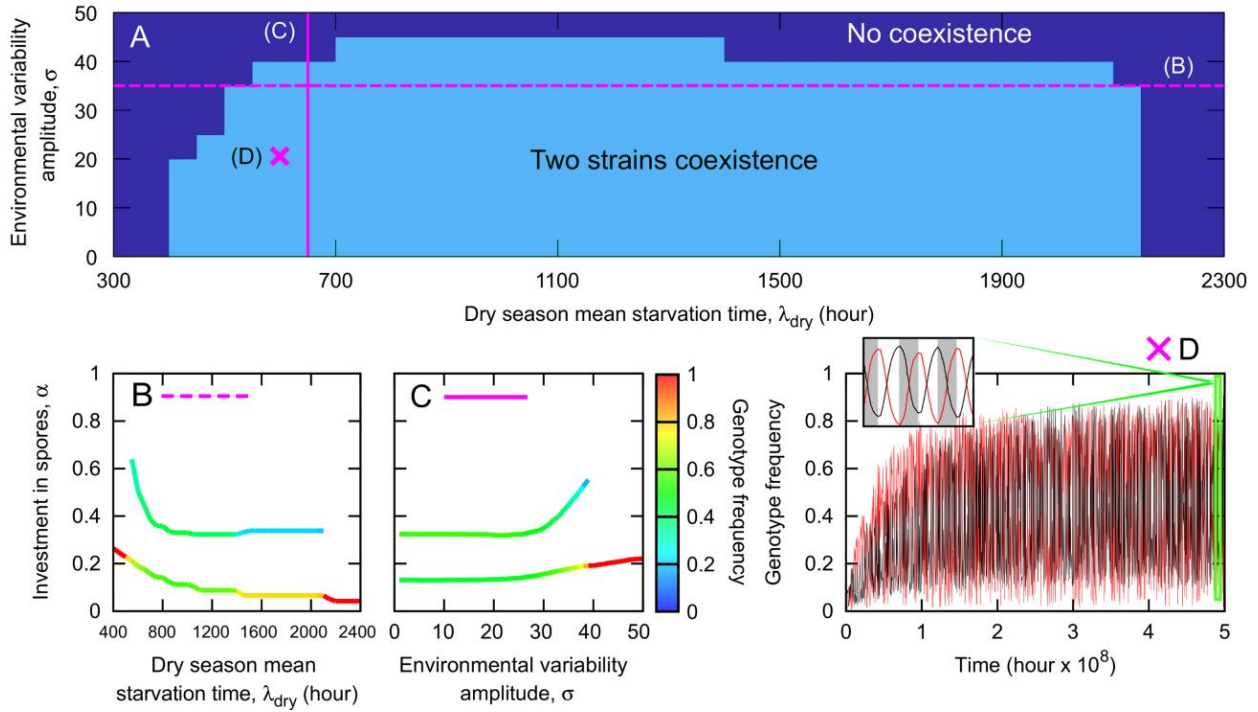

**Fig 6.** A milder dry season with resources replenishment enables strain coexistence. Parameters: $T_{wet}$=6 months, $T_{dry}$=6 months, $\lambda_{wet}$=50 hours. A) Regions of coexistence in the parameter space defined by the amplitude of the environmental variability (taken to be equal in both seasons) and the mean starvation time in the dry season. The space is scanned with a resolution given by $\Delta\sigma$=5 and $\Delta\lambda_{dry}$=50. The dashed (solid) magenta line and the magenta cross indicate the regions of the parameter space used in panels B (C) and D respectively. B) Surviving genotypes as a function of the dry season mean starvation time for a fixed amplitude of the environmental variability, $\sigma$=35. The color of the line changes according to the frequency of each genotype. C) Same as B but keeping $\lambda_{dry}$=650 hours constant and varying $\sigma$. D) Time series of the two coexisting genotypes, $\alpha$=0.142 (red) and $\alpha$=0.324 (black). The inset shows a zoom on the interval marked by the green window, gray time intervals correspond to the wet season and white intervals to the dry season. Surviving strains and their frequencies are obtained following the protocols explained in Materials and Methods.

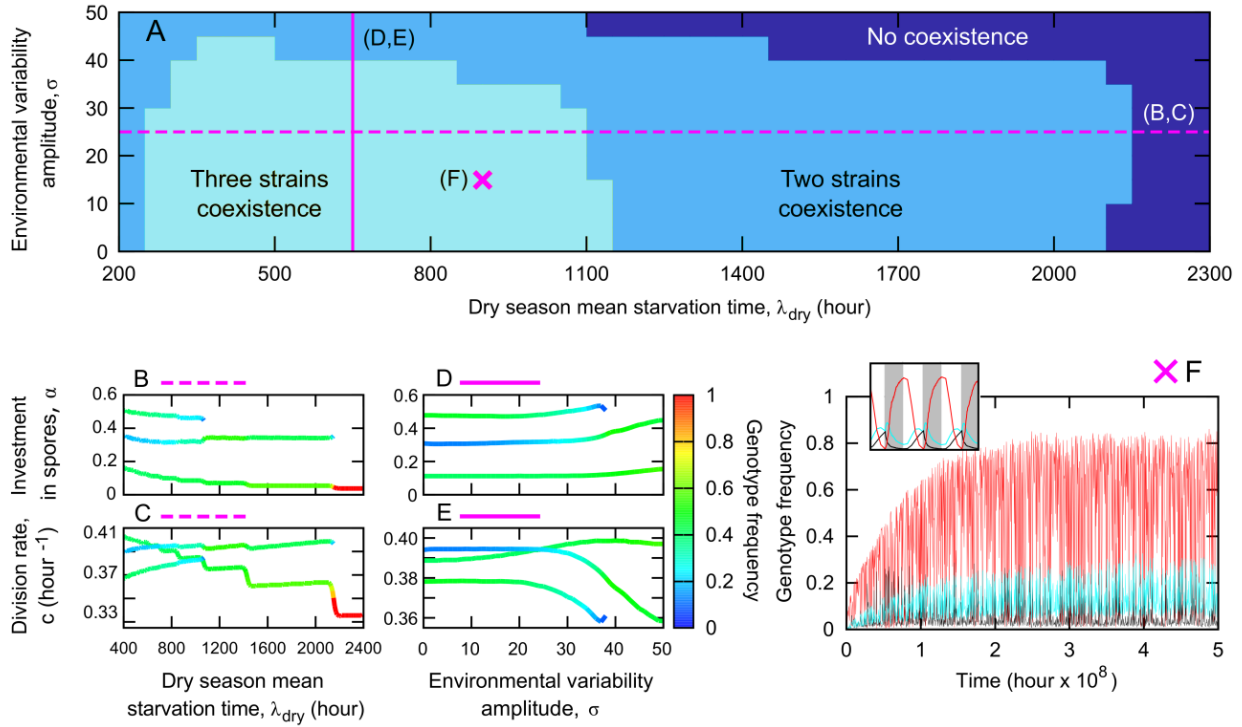

**Fig 7.** Strain diversity increases when multiple life-history tradeoffs are included. Parameters: $T_{wet}$=6 month, $T_{dry}$=6 month, $\lambda_{wet}$=50 hour. A) Regions of coexistence in the parameter space defined by the amplitude of the environmental variability (taken to be the same in both seasons) and the mean starvation time in the dry season. The space is scanned with a resolution given by $\Delta\sigma$=5 and $\Delta\lambda_{dry}$=50. The dashed (solid) magenta line and the magenta cross indicate the region of the parameter space used in panels B, C (D, E) and F respectively. B, C) Surviving genotypes defined by $\alpha$ (B) and $c$ (C) as a function of the dry season mean starvation time for a fixed environmental variability amplitude, $\sigma$=25. The color of the line changes according to the frequency of each genotype. D, E) Same as B and C but keeping $\lambda_{dry}$=650 hour constant and varying $\sigma$. F) Time series of the three coexisting genotypes: $\alpha$=0.085, $c$=0.380 (red); $\alpha$=0.324, $c$=0.390 (black) and $\alpha$=0.488, $c$=0.380 (cyan). The inset shows a zoom on the interval marked by the green window, gray time intervals correspond to the wet season and white intervals to the dry season. Surviving strains and their frequencies are obtained following the protocols explained in Materials and Methods.

| Exp. observation | Experimental trait involved | Model hypothesized trait | Hypothesized tradeoff | Functional consequence |
|---|---|---|---|---|
| Non-aggregated cells exist and are viable [5], [6] | - | Aggregator to non-aggregator ratio $\alpha$ | Non-aggregator vs. spore production | - Spores: resist starvation, delayed reproduction if food returns<br><br>- Non-agg.: less resistant to starvation, reproduce readily if food returns |
| Genotypes with more spores have smaller less viable spores [4] | Spore production | Division rate, $c$ | Spore production proxy for reproduction | Spore production correlates with $c$ |
| | Spore size | | Growth vs reproduction | Cell/spore size anti-correlates with $c$ |
| | Spore viability | | Survival vs reproduction | Spore viability anti-correlates with $c$ |

Table 1. Summary of the model. Starting from experimental observations, we build a socially-neutral model (no social interactions between genotypes) of cellular slime molds that incorporates measured life-history traits and tradeoffs via two hypothesized traits: the aggregator to non-aggregator ratio and the division rate.

| Description | Parameter | Value | Units |
|---|---|---|---|
| Rate of decrease of the survival probability | $\mu$ | $2 \times 10^{-3}$ | hour$^{-1}$ |
| Maximum lifetime of a vegetative cell | $T_{sur}$ | 200 | hour |
| Spore germination time | $\tau$ | 4 | hour |
| Spore mortality rate | $\delta$ | $2 \times 10^{-4}$ | hour$^{-1}$ |
| Fraction of aggregators that become spores | $s$ | 0.8 | --- |
| Food pulse size | $R_0$ | $10^8$ | # cells |
| Half-saturation constant of resources consumption | $R_{1/2}$ | $0.1 R_0$ | # cells |
| Resources exhaustion threshold | $R^*$ | 1 | # cells |
| Investment in aggregation | $\alpha$ | varied | --- |
| Division rate | $c$ | 0.173 or varied | hour$^{-1}$ |
| Scale factor of resources per division with cell size | $Q$ | $c_{max}/c$ | --- |
| Fastest division rate | $c_{max}$ | 0.45 | hour$^{-1}$ |
| Spore viability | $v$ | $1.1 - 2c$ | --- |
| Resistance to starvation parameter | $\beta$ | $3.1 - 4c$ | --- |

Table 2. Definition of the parameters and their values. The lower part of the table includes the traits that are allowed to evolve in some of the sections and their functional consequences.